\documentclass[useAMS,usenatbib,usegraphicx]{mn2e}
\usepackage{times} 

\newcommand{\simlt}{\lower.5ex\hbox{$\; \buildrel < \over \sim \;$}}
\newcommand{\simgt}{\lower.5ex\hbox{$\; \buildrel > \over \sim \;$}}
\newcommand{\vrot}{v_{\rm MAX}}
\newcommand{\be}{\begin{equation}}
\newcommand{\ba}{\begin{eqnarray}}
\newcommand{\ee}{\end{equation}}
\newcommand{\ea}{\end{eqnarray}}

\title[The SFH of intermediate $z$ late-type galaxies]
        {The Star Formation History of intermediate redshift late-type galaxies}
\author[I.~Ferreras, J.~Silk, A. B\"ohm, B. Ziegler]
{Ignacio Ferreras$^{1,2}$\thanks{E-mail:
ferreras@phys.ethz.ch}, Joseph Silk$^1$, Asmus B\"ohm$^3$, Bodo Ziegler$^3$\\
$^1$Physics Dept. Denys Wilkinson Building, Keble Road, Oxford OX1 3RH\\
$^2$Institut f\"ur Astronomie, Swiss Federal Institute of Technology
(ETH), H\"onggerberg Campus, HPF D8, CH-8093 Z\"urich, Switzerland\\
$^3$Universit\"atssternwarte G\"ottingen, Geismarlandstrasse 11, 
  D-37083 G\"ottingen, Germany}
\begin{document}

\date{Accepted for publication in MNRAS, August 9, 2004}

\pagerange{\pageref{firstpage}--\pageref{lastpage}} \pubyear{2004}

\maketitle

\label{firstpage}

\begin{abstract}
We combine the latest observations of disk galaxy photometry 
and rotation curves at
moderate redshift from the FORS Deep Field (FDF)
with simple models of chemical enrichment. Our method
describes the buildup of the stellar component through
infall of gas and allows for gas and metal outflows. In this
framework, we keep a minimum number of constraints and we 
search a large volume of parameter space, looking for the
models which best reproduce the photometric observations 
in the observed redshift range ($0.5<z<1$).
We find the star formation efficiency to correlate well with 
$\vrot$  so that massive disks are more efficient in the formation
of stars and have a smaller spread in stellar ages. This trend presents
a break at around $\vrot\sim 140$~km~s$^{-1}$. Galaxies
on either side of this threshold have significantly different
age distributions. This break has been already suggested by several
authors in connection with the contribution from either gravitational
instabilities or supernova-driven turbulence to star formation.
The gas infall timescale and gas outflows also present a correlation
with galaxy mass, so that massive disks have shorter infall timescales
and smaller outflow fractions.
The model presented in this paper suggests massive disks have formation
histories resembling those of early-type galaxies, with highly efficient
and short-lived bursts, in contrast with low-mass disks, which  
have a more extended star formation history. The ages correlate well
with galaxy mass or luminosity, and the predicted gas-phase metallicities
are consistent with the observations of local and moderate redshift
galaxies. One option to explain the observed shallow slope of the 
Tully-Fisher relation at intermediate redshift could be small episodes 
of star formation in low-mass disks.
\end{abstract}

\begin{keywords}
galaxies: spiral -- galaxies: evolution -- galaxies: formation -- 
galaxies: stellar content.
\end{keywords}

\section{Introduction}
Quantifying the star formation history of galaxies constitutes
one of the major unsolved issues towards a complete understanding of galaxy
formation. Within the current paradigm of structure formation, disk galaxies
represent the building blocks from which the whole zoo of galaxies
is formed. A detailed analysis of galaxy formation histories is hindered
by the complexity of star formation. By coupling simple assumptions 
about star formation and chemical enrichment to population synthesis models which
track the evolution of well-defined stellar populations, one can
constrain the possible scenarios of star formation in a semi-quantitative
way. Furthermore, by exploring galaxies at high redshift one 
takes advantage of the lookback times which allow us to probe back into
the past histories of these galaxies.

Disk galaxies have been  complicated  to analyze, mostly because of 
their ongoing star formation, dust and complex geometry which prevent us 
from getting a clear picture of their underlying stellar populations.
Qualitatively, the photo-spectroscopic properties of disk galaxies
such as our own Milky Way are compatible with a weak and continuous 
star formation rate (SFR) along with a stronger early episode of 
star formation whose amplitude correlates well with galaxy
type, being stronger in disks of earlier types (e.g. Kennicutt 1998a).

Studies of disks at moderate and high redshift seem to yield a
similar formation scenario.
Lilly et al. (1998) observed a sample of $341$ galaxies selected
from the CFRS and the LDSS redshift surveys, out to $z\simlt 1$,
finding bluer $U-V$ colours and stronger [OII] emission
at moderate redshift, which suggests an increase in the SFR
by a factor of about $3$ from local galaxies out to $z\sim 0.7$. 
However, the size function was not found to change with redshift, and
the larger changes are associated with the smaller galaxies
(however, see Giallongo et al. 2000; Vogt 2000).
This is indicative of earlier formation of massive disks, in agreement
with the lower gas fractions found in these 
systems (e.g. Bell \& de~Jong 2000).
Boissier \& Prantzos (2001) presented a detailed model of chemical 
enrichment in disk galaxies calibrated to the Milky Way
(Boissier \& Prantzos 1999). The application of this model to
disk galaxies at moderate redshift showed that large disks have
already completed their evolution by $z\sim 1$, whereas 
low-mass disks undergo a later evolution, resulting in a 
steepening of the slope of the Tully-Fisher relation with redshift.
A similar conclusion is reached by Ferreras \& Silk (2001)
when comparing sets of simple chemical enrichment models over 
a wide range of parameter space using the $z=0$ optical and NIR 
colour--$\vrot$ relation as a constraint. 
The star formation efficiency was found to correlate with 
galaxy mass, a point which is questioned by Boissier et al. (2001).
However, this controversy hinges on the degeneracy between
star formation efficiency and infall timescale. Both groups 
agree on the need for a rapid buildup of the stellar component
in massive disks, a property that can be achieved either by a
high efficiency of star formation or by a shorter infall timescale.

The evolution of the Tully-Fisher relation with redshift is 
a powerful discriminator of star formation histories.
Vogt (2000) presented a sample of $100$ faint disk 
galaxies in the range $0.2<z<1$ and 
no change was found in either shape or slope with respect to  local samples.
A zero point offset $<0.2-0.3$~mag was found in the rest frame $B$ band.
Milvang-Jensen et al. (2003) explored cluster/field differences
in the TF relation at moderate redshift on a smaller sample of 
$8$ disk galaxies in cluster MS1054-03 ($z=0.83$) as 
well as $19$ field spirals in the $z=0.15-0.9$ range. At fixed
rotation velocity, they found cluster spirals to be $\sim 0.5-1$~mag
brighter. However, their sample is too small to draw a firm
conclusion on the evolution of the Tully-Fisher relation. 
In a larger sample of disk galaxies in 3 clusters at moderate 
redshift ($z\sim 0.5$) Ziegler et al. (2003) did not find 
significant differences with respect to field disks.

\begin{figure}
\includegraphics[width=3.5in]{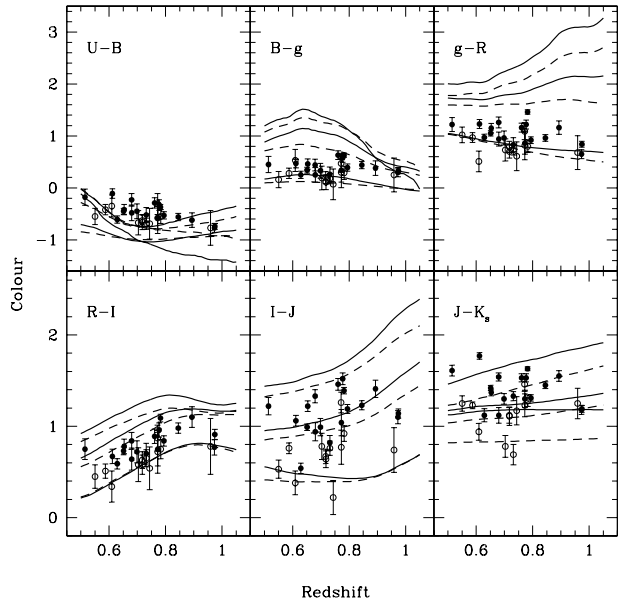}
\caption{Colour-redshift plots of the sample of FDF moderate redshift disk
galaxies used in this paper. The filled and empty dots represent
galaxies with $\log\vrot >2.2$ and $\leq 2.2$, respectively. The lines 
correspond to the colours of simple stellar populations from the latest 
models of Bruzual \& Charlot (2003), for three ages (from top to
bottom): 8, 1, and 0.1~Gyr. For each age two metallicities are 
considered: $Z_\odot$ (solid) and $Z_\odot /3$ (dashed). Notice the
bluer colours -- which are most sensitive to young stars -- suggest
young stellar ages, whereas the red/NIR colours imply older ages.
}
\label{fig:CCDs}
\end{figure}

The work on $113$ disk galaxies in the range $0.1<z<1.0$ from the 
FORS Deep Field on which this work is based (B\"ohm et al. 2004)
revealed a striking result, namely a shallow slope of the
rest frame $B$ band Tully-Fisher relation with respect to local disks.
Slowly rotating disks ($\vrot\simlt 100$~km~s$^{-1}$) at moderate redshift 
appear up to $2$~magnitudes brighter than their low-redshift counterparts.
One should question whether the samples at low and moderate
redshift correspond to a simple one-to-one mapping. Indeed, the sample
of Lilly et al. (1998) presents a large fraction of bright and small
disks at $z\sim 1$ which do not have local counterparts.

In this paper, we combine photometric data from the FDF disk sample 
with a model which convolves galactic chemical evolution histories
with stellar population synthesis in order to relate the photometric 
observations with a small set of phenomenological parameters which 
describe the major drivers of star formation in galaxies. 
We describe the observations in \S2 and our model in \S3. The results 
are discussed in \S4 and the resulting ages and metallicities are
presented in \S5. Finally, in \S6 we discuss our results and give 
the conclusions. Throughout this paper we use a $\Lambda$CDM cosmology
with $\Omega_m = 0.3$ and $H_0=70$~km~s$^{-1}$~Mpc$^{-1}$,
as suggested by the analysis of the angular power
spectrum of the Cosmic Microwave Background observed
by WMAP (Spergel et al. 2003). For this cosmology, the
age of the Universe at $z=0$ and $z=1$ corresponds to
$13.5$ and $5.8$~Gyr, respectively.

\section{Observations}
We use a subsample of the data set presented by Ziegler et al. (2002)
and B\"ohm et al. (2004) 
to study the redshift evolution of the Tully-Fisher relation. The
galaxies are extracted from the FORS Deep Field (FDF), a deep 
$(UBgRIJK_s)$ survey over a $6^\prime\times 6^\prime$ area 
(Appenzeller et al. 2000). 
Spectroscopy with a resolution $R\sim 1200$ was carried out 
at FORS$1+2$ mounted on the VLT. Throughout this paper we refer
to the intrinsic rotation velocities as $\vrot$ since these have
been obtained from the flat part of the rotation curves.
From the complete sample of $77$ galaxies with measured $\vrot$, we
selected those at redshifts $z\geq 0.5$ resulting in a 
sample of $30$ moderate redshift late-type galaxies, which 
cover a range of rotation velocity between $85$ and $400$~km~s$^{-1}$.
Figure~\ref{fig:CCDs} shows the colour-redshift diagrams of the $z>0.5$ 
sample of FDF galaxies used in this paper. The photometry is performed
over a 2~arcsec aperture, which corresponds to a projected physical size of 
$12-16$~kpc in the $z=0.5-1$ redshift range. The magnitudes were corrected for
galactic extinction, following Heidt et al. (2003) and intrinsic 
rest-frame extinction following Tully \& Fouqu\'e (1985), with the 
convention of a non-negligible extinction for face-on disks, so that
the minimum value, e.g., in rest-frame $B$-band is $A_B=0.27$~mag. 
For more details on the sample reduction and $\vrot$ determination 
see B\"ohm et al. (2004). In the appendix we have
explored the effect that a different dust correction would have on the
model predictions. The lines in figure~\ref{fig:CCDs}
show the  colours for a set
of simple stellar populations from the latest Bruzual \& Charlot (2003) 
models for two metallicities: solar (solid lines) and $Z_\odot /3$ (dashed 
lines), assuming three different ages (from top to bottom): 8, 1, and 0.1~Gyr.
These are predictions for simple populations with a single age and metallicity.
Notice that the bluer colours -- which are very sensitive to young stars --
suggest young ages, whereas redder colours are more compatible with older ages. 
A more realistic composite model will take into account the different 
contributions from various ages and metallicities to each passband.

\begin{figure}
\includegraphics[width=3.5in]{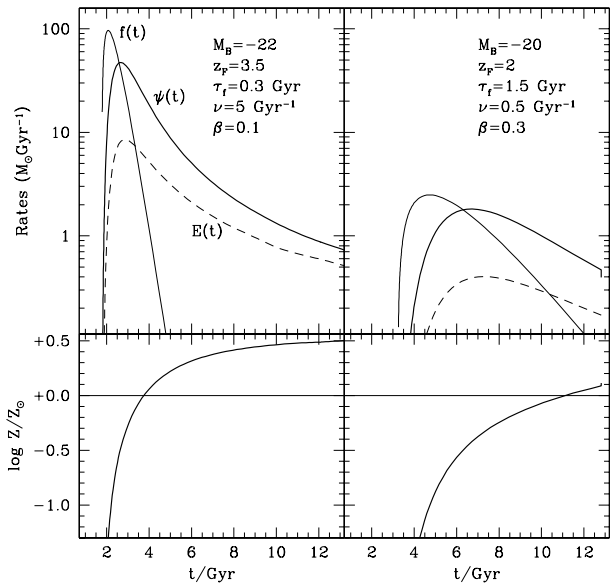}
\caption{A sample of the generic star formation histories 
used in the modelling throughout this paper. The SFHs shown here 
give good fits to the colours of a massive ($M_B=-22$; {\sl left panel}) 
and a lower-mass disk ($M_B=-20$; {\sl right panel}). The top panels show 
the star formation rate ($\psi(t)$), which is
driven by a Schmidt-type law fuelled by the infall of primordial
gas ($f(t)$). The dashed line ($E(t)$) shows the contribution to the
gas component from stars as they reach their endpoints of stellar
evolution. The lower panels
give the evolution of the metallicity with age. The parameters which correspond
to each SFH are labelled in the top panels (see text for details).
The mass rates are normalized to a final stellar mass content 
of $10^{11}$ (left) and $10^{10} M_\odot$ (right).}
\label{fig:sfh}
\end{figure}

\section{Modelling the SFH of disk galaxies}
We consider a phenomenological scenario of star formation and 
chemical enrichment based on the models presented in 
Ferreras \& Silk (2001). The rationale of our approach
is to try to minimize the number of parameters describing the
buildup of the stellar component. This enables us to run a fairly
big number of realizations of these models, scanning a large volume of
parameter space. We reduce the description of the process of star 
formation in galaxies to a few parameters, namely:
\begin{itemize}
\item[$\bullet$] {\bf Star Formation Efficiency:} Star formation is
  assumed to follow a variant of the Schmidt law (Schmidt 1959): 
  \be 
  \psi (t)=\nu\rho_{\rm TOT}\Big(\frac{\rho_g(t)}{\rho_{\rm TOT}}\Big)^n, 
  \label{eq:schmidt} 
  \ee 
  where $\rho_g$ is the gas volume density and $\nu$ is a parameter 
  describing the star formation efficiency, which is an inverse
  timescale for the processing of gas into stars. $\rho_{\rm TOT}$ is
  the total baryon density, which -- for a single zone model -- is 
  just a fixed number throughout the analysis. The only purpose of
  including it in equation~\ref{eq:schmidt} is to give $\nu$ the
  units of an inverse timescale.
  Based on the best fit to observations from a local sample of
  normal spiral galaxies (Kennicutt 1998b) we decided to use $n=1.5$. 
  This exponent is also what
  one obtains for a star formation law that varies linearly with 
  gas density and inversely with  the local dynamical time: 
  $\psi\propto\rho_g/t_{dyn}=\rho_g\sqrt{G\rho_g}\propto\rho_g^{1.5}$.
\item[$\bullet$] {\bf Infall:} The infall of primordial gas is described by
  a set of two parameters. We assume the infall rate -- $f(t)$ -- 
  to be a delayed exponential function with an infall timescale
  ($\tau_f$). We characterize the starting point at some epoch
  by a formation redshift $z_F$, so that the age of the Universe when
  the infall begins is $t(z_F)$. Defining $\Delta t\equiv t-t(z_F)$, we can
  write the infall rate as:
  \be
  f(t)\equiv\frac{\rho_{\rm TOT}}{\tau_f^2}\Delta t\, e^{-\Delta t/\tau_f}
  \ee
  This function rises quickly to its maximum value at $\Delta t=\tau_f$ and
  then declines on a slower timescale. After a time $\Delta t_{90}=3.9\tau_f$
  90\% of the total baryon content ($\rho_{\rm TOT}$) has fallen on the galaxy.
  Notice that this choice of infall rate allows for a very wide of
  scenarios. Besides the standard model of a quickly rising rate with a 
  slowly decaying branch, our parameter space also includes scenarios
  with an increasing infall rate -- i.e. when long timescales 
  ($\tau_f\sim 4-5$~Gyr) and late formation epochs ($z_F=2$) are chosen.
  It is only the data that eventually constrain these scenarios.
\item[$\bullet$] {\bf Outflow fraction:} The ejection of gas and metals 
  from winds triggered 
  by supernova explosions constitutes another important factor 
  contributing to the final metallicity of the galaxy. We define 
  a parameter ($0\leq\beta\leq 1$) which represents the fraction of gas and
  metals from stellar ejecta and lost by the galaxy. Even though one could
  estimate a -- model-dependent -- value of $\beta$ given the star 
  formation rate and 
  the potential well given by the total mass of the galaxy
  (e.g. Larson 1974, Arimoto \& Yoshii 1987), we leave $\beta$
  as a free parameter, only to be constrained {\sl a posteriori} 
  by the data.
\end{itemize}
The stellar yields are simplified in the model by a truncated power law 
fitting the results from Van~den~Hoek \& Groenewegen (2002)
for Intermediate Mass Stars (IMS) (i.e. $M\simlt 8M_\odot$). 
The yields from massive stars -- which undergo core-collapse -- are 
obtained from Thielemann, Nomoto \& Hashimoto (1996) assuming 
solar metallicity progenitors. 
A Salpeter (1955) IMF was used in the $0.1-60 M_\odot$ 
mass range. Other IMFs such as Chabrier (2003) were tested, to find no
siginificant difference. After all, the slope of the IMF in the upper
mass region -- which mostly controls the chemical enrichment -- 
is similar, and the colours do not vary significantly. The most important
difference between these two IMFs is a systematic change in the
stellar mass-to-light ratios. We refer the reader to Ferreras \& Silk (2001) 
for details of the model applied to disk galaxies.

Figure~\ref{fig:sfh} shows two sample star formation histories obtained
by this parametrization. It corresponds to the best fits obtained for a
typical massive ({\sl left}) and low-mass disk ({\sl right}). The parameters
used for each one are given in the top panels.
The star formation rate ($\psi$), infall rate ($f$) and stellar ejecta
($E$) are shown, normalized to a final stellar mass content at $z=0$
of $10^{11}$ ({\sl left}) and $10^{10}M_\odot$ ({\sl right}). 
The bottom panels give the evolution of the gas phase metallicity.
Several models were run with formation redshifts between $2$ and $5$
and the results vary smoothly between these two extrema, which 
correspond to a formation time of $3.2$ and $1.2$~Gyr, 
respectively. More realistic models should have a well defined
correlation between this formation redshift and the mass of the
galaxy. However, in our spirit of keeping the number of constraints
to a minimum, we ran six grids of models for each galaxy with various 
values of $z_F$ and chose the one with the minimum $\chi^2$.
We want to emphasize that the trends presented in this paper are
fairly robust and insensitive to the particular choice of $z_F$.
The remaining three parameters $(\nu ,\beta ,\tau_f)$ are explored 
over a large volume of parameter space as described in more detail below. 

\subsection{Data constraints}
For each choice of parameters ($\tau_f,\nu,\beta$) we solve
the chemical enrichment equations which give us a star formation
history that is subsequently convolved with the latest
population synthesis models of Bruzual \& Charlot (2003) in order
to generate a spectral energy distribution (sed) at
the observed redshift of the galaxy under scrutiny ($0.5<z<1$).
A $\chi^2$ function is constructed by comparing the model prediction
with the 2~arcsec aperture photometry.
We define the $\chi^2$ function as:
\be
\chi_z^2\equiv\mathop{\sum}_{i=1}^6
\frac{[{\cal C}_z(i)-{\cal C}_z^m(i)]^2}{\sigma_z(i)^2},
\ee
where ${\cal C}_z(i)$ is the observed colour within a 2~arcsec
aperture -- which represents a physical projected distance between 
$12$ and $16$~kpc in the redshift range $0.5<z<1$. The fluxes were 
obtained from the coadded FDF images after convolving them to 
the same seeing -- see Heidt et al. (2003) for details. The indices
$i=\{1,2,3,4,5,6\}$ correspond to colours $U-B$, $B-g$, $g-R$, $R-I$, 
$I-J$ and $J-K_s$, respectively. The photometric uncertainties in 
the colours are $\sigma_z(i)$ and are typically in the range $0.05-0.2$~mag. 
The model 
prediction in each filter is given by ${\cal C}_z^m(i)$. 
For each model realization we compute the rest-frame $B$-band absolute
luminosity at the redshift of the galaxy. These luminosities are
normalized with respect to the observed apparent brightness in a
passband closest to the rest-frame $B$ band. In our sample this 
corresponds to $R$ for $z<0.85$ and $I$ for $z\geq 0.85$ (B\"ohm et al. 2004). 
The total apparent magnitudes correspond to the MAG\_AUTO algorithm of the 
Source Extractor package (Bertin \& Arnouts 1996).

\subsection{Finding the best fit}
Searching for the choice of parameters which give the best fit
according to the $\chi^2$ function defined in the previous section is
a rather computer-intensive endeavour. We decided to explore a set of 
$6$ formation redshifts, namely $z_F=\{ 2, 2.5, 3, 3.5, 4, 5\}$.
The remaining three parameters, namely $(\nu ,\tau_f,\beta )$ 
are searched for each galaxy in our sample in a $32\times 32\times 32$ grid 
over a wide range of values, namely:
\begin{center}
$-1.3\leq\log(\nu /{\rm Gyr}^{-1})\leq +1.5$,\\
$\,\,0\leq\beta\leq1$,\\
$-1.3\leq\log(\tau_f/{\rm Gyr})\leq+0.7$.\\
\end{center}
The best fit to the data is determined by
the maximization of the likelihood function 
${\cal L}(z_F, \nu ,\beta ,\tau_f)\propto\exp [-\frac{1}{2}\chi^2(z_F, 
\nu ,\beta ,\tau_f)]$, 
where $\chi^2(z_F, \nu ,\beta ,\tau_f)$ is described above. 
With the uncertainties inherent to any modelling of
galactic chemical enrichment and given that the constraints used in this 
paper are only based on a few photometric measurements, we caution the 
reader to take these results as semi-quantitative trends. The strongest
value of the model predictions lie in the {\sl relative} difference between
parameters among galaxies. In the following
figures we show the best fits for each galaxy along with the $1\sigma$
confidence level obtained by the likelihood function.

\begin{figure}
\includegraphics[width=3.5in]{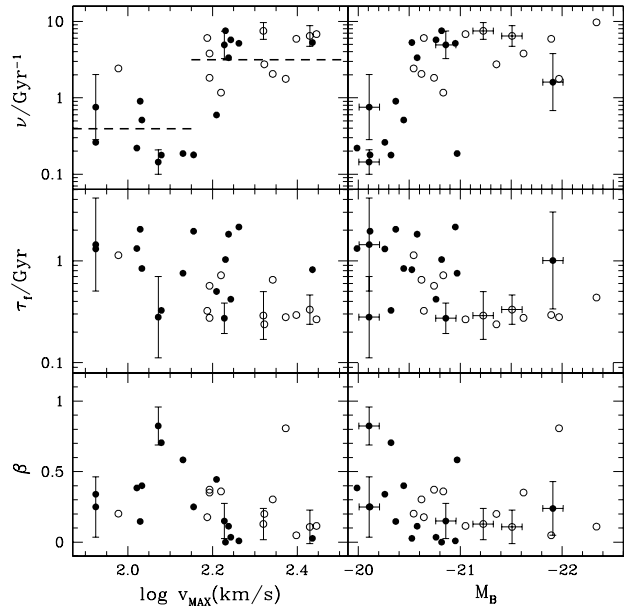}
\caption{Predicted model parameters as a function of rotation
velocity ({\sl left}) or absolute $B$-band luminosity ({\sl right})
for the complete ($z>0.5$) FDF moderate redshift sample. 
The filled and hollow circles represent the $z<0.75$ 
and $z\geq 0.75$ subsamples, respectively. 
Typical marginalized $1\sigma$ error bars are shown.
The parameters are (from top to bottom) star formation efficiency ($\nu$), 
infall timescale ($\tau_f$), and outflow fraction ($\beta$). 
The dashed lines in the top panels represent the median value of
the star formation efficiency, separating the sample at 
$\vrot\sim 140$~km~s$^{-1}$.
}
\label{fig:params}
\end{figure}

\section{The star formation history of disk galaxies}
Figure~\ref{fig:params} shows the prediction for the best fit values of the
parameters controlling the star formation history as a function of
rotation velocity ({\sl left}) or $B$-band luminosity ({\sl right}). 
The sample
is separated into a lower redshift ($0.5\leq z<0.75$; filled circles) and
a higher redshift sub-sample ($z\geq 0.75$; hollow circles). Some typical
($1\sigma$) marginalised error bars are shown. Even though the uncertainties
are rather large -- as expected from any analysis based on broadband photometry
-- one can still hint at a significant correlation between the star formation
efficiency ($\nu$; top panels) and rotation velocity. The dashed lines give the
median of $\nu$ separating the sample at $\vrot\sim 140$~km~s$^{-1}$. 
A clear increase in the star formation efficiency is seen for the more
massive disks.
Furthermore, the position of this ``break'' is reminiscent 
of the recent results 
of Kauffmann et al. (2003) on a large sample of $10^5$~galaxies from the
SDSS. Their study showed a clear bimodality in the distribution of 
age-sensitive observables such as the 4000\AA\ break strength 
or H$\delta_A$ Balmer absorption at a stellar mass 
$M_\star\sim 3\times 10^{10}M_\odot$, which corresponds to 
$\log\vrot\sim 2.2$ for a reasonable choice of stellar $M/L$. 
The infall timescale (middle panels) appears also to be correlated 
with rotation velocity, so that low-mass disks seem to have longer infall
timescales. The data suggests that $\nu$ and $\tau_f$ present similar
correlations, a well-known degeneracy (e.g. Ferreras \& Silk 2003) 
which would require more detailed photo-spectroscopic observables 
to disentangle. Since this analysis is based on broadband photometry
which is very much affected by the infamous age-metallicity degeneracy,
we decided to test the robustness of our models with respect to a
possible systematic effect from the dust correction applied. The appendix
compares two possible dust corrections and proves that the bimodality
obtained in the star formation efficiency is fairly robust to these
corrections.

\begin{figure}
\includegraphics[width=3.5in]{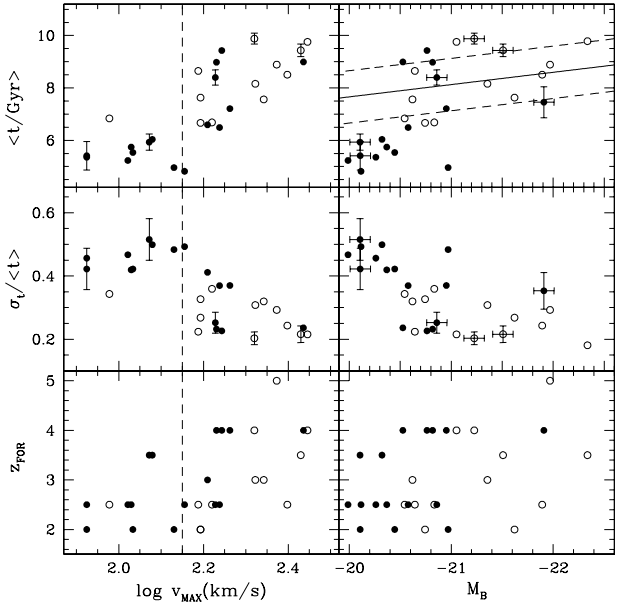}
\caption{Average ({\sl top}) and RMS ({\sl middle})
values of the best fit stellar age distribution 
predicted for each galaxy {\sl after being evolved to $z=0$}. 
The vertical dashed line gives 
the position of the break at $\vrot =140$~km~s$^{-1}$.
Notice that galaxies with larger rotation velocities than 
this break present a larger scatter towards older ages.
The solid line in the top-right panel gives the results 
from Bell \& de~Jong (2000) along with a rough estimate of the scatter
(dashed lines).
The bottom panels show the formation redshift which gives
the best fit (see text for details).
\label{fig:agemet1}}
\end{figure}

At this point we should mention the result of Boissier et al. (2001) who 
claim the efficiency to be uncorrelated with galaxy mass. However, 
their definition of efficiency is different from ours: 
$\epsilon\equiv\psi /M_g$, where $M_g$ is the total gas mass. We can 
translate between these two definitions of efficiency:
\be
\nu = \epsilon \Big( \frac{M_g}{M_{\rm TOT}}\Big) ^{-0.5}.
\label{eq:eff}
\ee
The total baryon mass in the system is: $M_{\rm TOT}=M_g+M_\star+M_{ej}$, 
where the last term refers to the gas mass ejected from stars out of the galaxy.
This term can be written as roughly: $M_{ej}\sim \beta R M_\star$,
where $\beta$ is the outflow fraction defined in \S3 and $R$ is 
the returned fraction, i.e. the fraction of gas ejected from stars
(see e.g. Tinsley 1980). If we write the observed gas fraction as
$f_g=M_g/(M_g+M_\star)$, then we can rewrite equation~\ref{eq:eff} as:
\be
\nu = \epsilon \Big[ \frac{1}{f_g}(1+\beta R) -\beta R\Big]^{0.5}
\sim \epsilon f_g^{-0.5}.
\label{eq:eff2}
\ee
The returned fraction is in the range $0.2-0.3$ for most IMFs, so that
we can roughly neglect terms $O(\beta R)$ and get the scaling
only with respect to the gas fraction. 
Therefore, our results are compatible with the findings of 
Boissier et al. (2001) as long as the gas fraction decreases with 
galaxy mass. Bell \& de~Jong (2000) find a 
correlation between the gas
fraction and $K$ band absolute luminosity, namely 
$f_g=0.8+0.14(M_K-20)$, which -- assuming a constant $\epsilon$ and 
using equation~\ref{eq:eff} -- approximately  gives a variation 
in $\log\nu$ of $+0.5$~dex between bright and
faint disks, roughly the same range as the one shown in figure~\ref{fig:params}
for subsamples in the same redshift range. Brinchmann \& Ellis (2000) 
studied the stellar masses and the star formation rates in a sample 
of $321$ field galaxies out to $z\sim 1$ and they found a clear 
decrease of the specific star formation rate 
-- defined as $\psi /M_\star$ -- with stellar mass ($M_\star$). 
Furthermore, $\psi /M_\star$ was found to increase with redshift.
We can write the specific SFR in terms of the efficiency:
\be
\frac{\psi}{M_\star}=\nu\frac{f_g^{1.5}}{1-f_g}\sim\epsilon\frac{f_g}{1-f_g}.
\label{eq:ssfr}
\ee
The anticorrelation found between $\psi /M_\star$ and stellar mass 
is thereby a signature of the decreasing gas fraction with galaxy mass. 
Notice that in figure~\ref{fig:params} the subsample with
the higher redshift (hollow circles) has a systematically higher 
star formation efficiency ($\nu$), which is to be expected from 
equation~\ref{eq:ssfr} as the gas fraction will increase with redshift
at a fixed rotation velocity.

\begin{figure}
\includegraphics[width=3.5in]{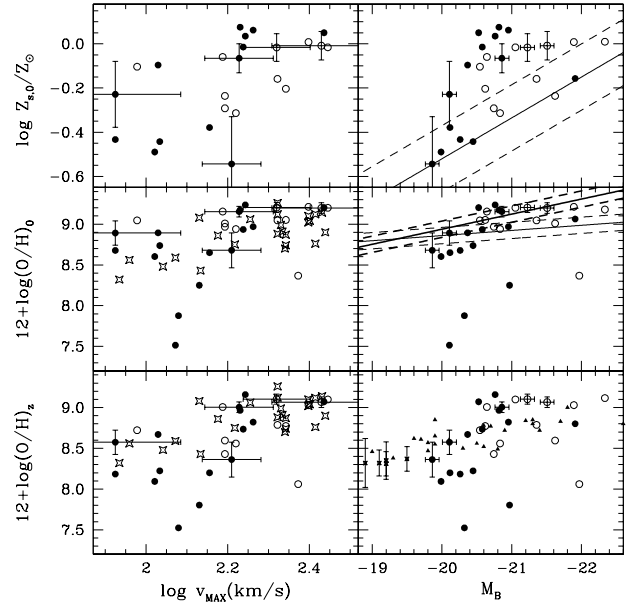}
\caption{Best average stellar ({\sl top}) and gas phase ({\sl middle, bottom}) 
metallicities of the sample.
The subscript $_0$ refers to metallicities measured after evolving the
best-fit SFH to zero redshift, whereas $_z$ refers to a measurement at the
observed redshift. The data are shown along with 
1-$\sigma$ error bars. 
The solid line in the top-right panel gives the results
from Bell \& de~Jong (2000) along with a rough estimate of the scatter
(dashed lines).
The stars in the left panels are oxygen gas
abundances from a sample of local disk galaxies (Garnett 2002),
The stars and triangles in the bottom-right panel give the gas-phase 
abundance measured in the $z>0.5$ subsample of moderate redshift 
galaxies from CADIS (Maier et al. 2004) and DGSS (Kobulnicky et al. 2003), 
respectively. 
The lines in the middle-right panel represent the luminosity-metallicity
relation from a sample of $\sim 53,000$ galaxies from SDSS
(thick lines; Tremonti et al. 2004) and from the local (K92+NFGS+KISS) 
sample explored in Kobulnicky et al. (thin lines; 2003 and references therein). 
A $\pm 0.1$~dex scatter is assumed.}
\label{fig:agemet2}
\end{figure}

The bottom panels of figure~\ref{fig:params} give the gas outflow fraction,
which weakly hints at an anticorrelation with rotation velocity.
Bright disks are redder and faint disks are bluer. 
This colour trend can be explained either by a range 
of metallicities --- in which the outflow fraction $\beta$ would play the main 
role --- or by a mixed range of ages and metallicities -- 
implying a stronger correlation of $\nu$ or $\tau_f$ with the 
luminosity of the galaxy. 
One can only estimate the weight of these
two factors by analyses such as the one presented in this
paper. However, the use of a few photometric data points 
to constrain the star formation history prevents us from extracting an
unambiguous conclusion and only allows us to give a semi-quantitative 
result. The current spectroscopic data from the FDF sample cannot be
used for these purposes. Indeed, the spectra from our $z>0.6$ galaxies only 
feature one emission line -- [OII]/3727\AA\ -- with a very low 
signal-to-noise ratio in the continuum.
A detailed spectroscopic analysis on a sample such as this one
will doubtlessly give us a clearer prediction of which mechanisms 
play a key role in the buildup of the stellar component in disk galaxies.

\section{Ages and Metallicities}
For every choice of star formation history as the one shown in 
figure~\ref{fig:sfh} one can compute the moments of the age and
metallicity distributions. Figure~\ref{fig:agemet1} presents the
average (top panels) and RMS (middle panels) ages of the stellar
component along with a few characteristic $1\sigma$ error bars. 
The filled and hollow circles correspond
to the lower ($0.5\leq z<0.75$) and higher redshift galaxies 
($z\geq 0.75$), respectively. 
Since the ages and metallicities will be strongly dependent on the
redshift of each galaxy, we decided to put all of them at the same
redshift. Hence, the values shown in figure~\ref{fig:agemet1} 
correspond to the stellar populations expected {\sl after evolving 
the predicted star formation histories to $z=0$}. The bottom panel
shows the formation redshift which give the best fit to the observed
colours. Higher formation redshifts seem to be favoured by massive
disks. The position of the break in the analysis
of the star formation efficiency is shown as a vertical dashed line. 
One sees a clear increase both in the star formation efficiency 
and the average age 
as a function of rotation velocity. The massive disks seem to have a
slightly smaller RMS scatter in the age, a result obviously generated
by their higher star formation efficiency.

Figure~\ref{fig:agemet2} presents the results for the average
mass-weighted stellar metallicity ({\sl top}) and the gas phase 
metallicity ({\sl middle}) of the best fit models evolved to zero redshift, 
and the gas phase metallicity at the observed redshift ({\sl bottom}). 
Notice that the stellar age and metallicity estimates presented
in this paper are in rough agreement with the results from the data
presented by
Bell \& de~Jong (2000). Our data does not allow us to
pursue an accurate analysis of the ages and metallicities 
as a function of central surface brightness, but we find a similar
correlation as Bell \& de~Jong with respect to absolute luminosity.
The solid line in the top-right panels of figures~\ref{fig:agemet1}
and \ref{fig:agemet2} show their fits as a function of absolute 
luminosity. The sample was originally presented as a function of
$M_K$ and we used their published $B-K$ colours to show it
with respect to $M_B$.
The dashed lines give a rough estimate of the large
scatter found in their sample. 
Gas metallicities are computed assuming a solar abundance ratio
in order to translate our metallicities (Z) into O/H measurements.
We assume a solar value of $12+\log(O/H)=8.7$ (Allende Prieto, Lambert \& 
Asplund 2001). A clear luminosity-metallicity relation can be seen, which is
consistent with local observations (K92+NFGS+KISS as referenced in 
Kobulnicky et al. 2003; thin lines in the middle panel) or with 
the recent analysis of the oxygen abundances
of a large sample ($\sim 53,000$) of galaxies from SDSS (Tremonti 
et al. 2004; thick lines). The dashed lines represent a $\pm 0.1$~dex 
scatter corresponding to typical error measurements. 
The triangles in the bottom-right panel are gas metallicity measurements
from the Groth Strip survey in the range $0.26<z<0.82$ 
(Kobulnicky et al. 2003) and the star symbols 
at $M_B>-20$ correspond to moderate redshift galaxies from 
CADIS (Maier, Meisenheimer \& Hippelein 2004). Our results are
also compatible with the local estimates of metallicity from the 
sample of Garnett (2002), shown as star symbols in the bottom-left panel 
with respect to rotation velocity.

\section{Discussion}
A simple phenomenological description of the star formation history allows
us to explore a large number of possible histories only to be constrained
by the observations. In this paper we have used the moderate redshift
sample of disk galaxies from the FORS Deep Field. Comparing the
aperture photometry in several passbands between data and simple
models of chemical enrichment across a wide spectral range ($U\cdots K_s$) 
we have found an increase of the 
star formation efficiency -- as defined in equation~\ref{eq:schmidt}  --
with rotation velocity. A clear threshold appears around 
$\vrot\sim 140$~km~s$^{-1}$
so that slowly rotating disks present uniformly 
low star formation efficiencies
and younger stellar populations with a larger spread of stellar ages.
More massive disks show a wide range of efficiencies.
A similar conclusion was reached by Ferreras \& Silk (2001) on the analysis
of the local sample of UMa cluster disks from Verheijen (2001). 
The thorough analysis of observational data across a wide spectral 
range -- from the radio to the UV -- probing the stellar and gas 
component of a sample of 928 
nearby disk galaxies led Gavazzi, Pierini \& Boselli (1996) to conclude
that, to first order, mass is the driving parameter of galaxy evolution.
They suggested that a positive correlation between star formation 
efficiency and galaxy mass should be invoked to explain the observations.
Furthermore, this trend is consistent with the analysis of 
Dalcanton et al. (2004) who also find a threshold in $\vrot$ when 
exploring the thickness of dust lanes in disks. 
They relate this threshold to a tradeoff between the 
two main causes of turbulence in the ISM,
namely supernova feedback and gravitational instabilities.
The former is the dominant factor in low-mass disks and 
produces a higher velocity dispersion in the gas, which results in 
lower star formation efficiencies. In contrast, Dalcanton et al. suggest 
turbulence in massive disks to be mainly caused by gravitational 
instabilities. The lower velocity dispersion implied by these instabilities 
results in an increased efficiency of star formation. Along similar lines,
Silk (2001) suggested a Schmidt-like star formation law:
\be
\psi\sim\Omega\rho_g\Big( \frac{\sigma_g}{v_{\rm SN}}\Big) ,
\ee
where $\Omega$ is the disk angular velocity, $\sigma_g$ is the cold
gas velocity dispersion and $v_{\rm SN}$ is a characteristic velocity
defined from supernova feedback as the specific momentum injected
by supernovae per unit star formation rate. In this framework, the 
results from figure~\ref{fig:params} suggest a major change in the
ratio $(\sigma_g/v_{\rm SN})$ at around $\vrot\sim 140$~km~s$^{-1}$

This bimodal trend has been explored in a large dataset
of $\sim 120,000$~galaxies drawn from the SDSS (Kauffmann et al. 2003). 
Low-mass galaxies were found to have
younger stellar populations and low concentrations, with some of these
systems having experienced recent star formation. At stellar masses above
$\sim 3\times 10^{10} M_\odot$, galaxies feature older populations and the
high concentrations typical of spheroidal systems. In the light of our
models, this spheroidal-like behaviour of large disks reveals the high
star formation efficiency as expected in bulges or elliptical galaxies.
Furthermore, a recent paper
(Kannappan 2004) has explored the gas-to-star mass ratios using $u-K$ 
colour as a proxy. The sample of $\sim 35,000$ galaxies
was extracted from
the SDSS and 2MASS databases and a similar bimodal trend was found, with
a threshold around a similar stellar mass, i.e. $2-3\times 10^{10}M_\odot$.
Our results confirm this bimodal trend in disk galaxies out to $z\sim 1$.

\begin{figure}
\includegraphics[width=3.5in]{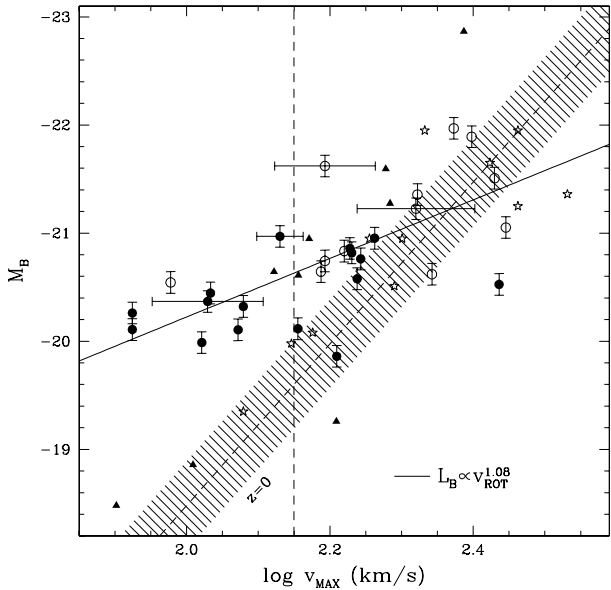}
\caption{The Tully-Fisher relation at moderate redshift. The 
filled and hollow circles correspond to the $z<0.75$ and 
$z\geq 0.75$ FDF subsamples, respectively. A few characteristic error bars 
in $\log\vrot$ are shown. The stars and triangles represent the 
($z\simgt 0.4$) data from Vogt et al. (1997) and the field 
disks from Milvang-Jensen et al. (2003), respectively. 
The solid line is a linear fit to the
FDF data. The fit to the local sample of Pierce \& Tully (1992) 
is also shown as a dashed line, with a shaded region representing
the scatter.
}
\label{fig:TFz}
\end{figure}

The star formation histories proposed in this paper for disk galaxies
allow us to determine
reliable evolution corrections and rest-frame luminosities. 
Figure~\ref{fig:TFz} shows the Tully-Fisher relation with respect to
rest-frame $M_B$. 
The model absolute luminosities are very similar to those already presented by
Ziegler et al. (2002) and B\"ohm et al. (2004) for which 
the $k$ corrections were obtained using the population
synthesis models of M\"oller et al. (2001).
As a comparison, we also show (at $z\simgt 0.4$) the field sample 
of Milvang-Jensen et al. (2003; triangles) as well as the sample from the 
DEEP Groth Strip Survey (Vogt et al. 1997; stars). 
The solid line is a fit to our data, whereas the local 
Tully-Fisher relation is shown as a dashed line and the scatter
is represented by the shaded region (Pierce \& Tully 1992).
The FDF sample clearly gives a shallower Tully-Fisher relation with respect
to the local sample. The small size and scatter in the sample of 
Milvang-Jensen et al. (2003) prevents us from making any estimate of their
mismatch with respect to the local TFR. Vogt et al. (1997) give a 
result which is more compatible with the observations at $z=0$. However, 
their selection favours large disks which biases the sample towards 
early-type systems. B\"ohm et al. (2004) suggested this slope change
could be caused  by a mass-dependent
luminosity evolution. In this paper we do not attempt
to relate the bright and slowly rotating disks from the FDF sample 
with local low-mass disk galaxies. In fact, a na\"\i ve one-to-one mapping
of local and high redshift Tully-Fisher relations would result in an
anticorrelation between star formation efficiency and galaxy mass.
We only use the aperture photometry of the FDF disk galaxies in order
to constrain their star formation histories. The rotation velocities
are then used to assign a ``size'' parameter to each galaxy. The evolution
of the slowly rotating disks from our analysis to $z=0$ results in 
a population of bright disks which are not seen in local surveys.
Hence, our results conclude that these objects do not have local
counterparts, in agreement with Lilly et al. (1998)
who find a population of small and bright disks at $z\simgt 0.7$ 
which do not have counterparts in the local galaxy census. 
A simple estimate of the contribution from young stars can
be done if we consider the combination of two simple stellar
populations, an old one which represents the underlying population
and a young one with a fractional {\sl mass} contribution $f_Y$
which corresponds to the recent episode of star formation. We can
then relate the actual luminosity of the galaxy ($L$) to the luminosity of
the underlying old stellar component ($L_O$), namely
\be
\frac{L}{L_O}=\frac{M\Upsilon^{-1}}{M_O\Upsilon_O^{-1}}=
1 + \Big( \frac{f_Y}{1-f_Y}\Big)\frac{\Upsilon_O}{\Upsilon_Y},
\ee
where $\Upsilon =M/L$ is the mass-to-light ratio, and the subscripts
$O$ and $Y$ denote the old and young component, respectively.
Let us consider a small fractional contribution, say $f_Y=0.1$.
Assuming a metallicity $\log (Z/Z_\odot )=-0.5$ for both components, 
and an age of $6$~Gyr for the old population -- corresponding to a
formation redshift $z_F\sim 5$ for a galaxy at $z=0.75$ --
this 10\% contribution in mass from young stars translates into 
$\Delta M_B = M_B - M_{B,O} = 2.5\log (L_{B,O}/L_B) = -1.9$~mag 
for a young component with $t_Y=0.1$~Gyr and $-0.95$~mag 
for $t_Y=0.5$~Myr. A couple of
Gyr after the episode has subsided (i.e. $t_O=8$~Gyr and $t_Y=2$~Gyr) 
the same 10\% contribution implies $\Delta M_B=-0.36$~mag.
Hence, small episodes of star formation could account for the
presence of these bright and slowly rotating disks.
We should mention that no strong evolution is expected in 
$\vrot$ from mass accretion from $z=1$ to the present time. 
In our adopted $\Lambda$CDM cosmology accretion is mostly 
suppressed at low redshift by the growth factor in linear 
theory. Indeed, the disk galaxy models of van~den~Bosch (2002)
show that the rotation curves do not change more than
$\sim 5$\% in this redshift range.

Nevertheless, this does not invalidate the correlation between $\vrot$ 
and the star formation efficiency, but rather, strengthens it. 
The progenitors of the local slowly rotating disks without any recent 
star formation will be below the detection
threshold of the FDF sample, which implies even lower efficiencies should
be considered at low $\vrot$, 
in agreement with the analysis of Ferreras \& Silk (2001).
Their prediction of a {\sl steepening} of the slope of the TFR with redshift
implies disks with $\vrot\simlt 100$~km~s$^{-1}$ at $z\sim 1$
should be fainter than $M_B\sim -18$, a result which is also consistent
with the models of Boissier \& Prantzos (2001).  It is intriguing that 
there is evidence  for very modest episodes of late star formation  
even in nearby early-type galaxies (Trager et al. 2000).
Theory strongly suggests that such episodes should increase with 
redshift. This is commonly accepted as plausible in galaxy compact  
groups and protoclusters, where the merger frequency is high. However a
comparably strong statement in the environs of typical disk galaxies
is puzzling. Very few dwarfs are found near the Milky Way or M31 relative to the
predictions of the commonly accepted CDM model. The usual explanation is 
that companion dwarfs have been stripped of gas and indeed of stars, as 
suggested by the evidence for tidal streams (Ferguson et al. 2002) 
and the ongoing disruption of the Sagittarius dwarf galaxy 
(e.g. Majewski et al. 2003). Such episodes must 
have been more frequent in the past if theory --  which predicts an order of 
magnitude more dwarfs than are observed -- is to be reconciled with the 
observations.  Dwarf disruption will include both compression 
of gas in the dwarf core as well as gas ejection into the gas 
reservoir of the host galaxy. Both effects will result in transient  
episodes of star formation. These enhanced phases of star formation, 
although individually modest and short-lived, could account for as much 
as half of the current stellar mass. Such minibursts are also inferred 
to have occurred in the Milky Way from  the past history of star formation 
inferred from analysis of chromospheric age indicators in the local 
stellar population (Rocha-Pinto et al. 2000). The inferred stages of episodic 
brightening at a lookback time of $5 $~Gyr or more could help account 
for the brightening that we infer of small disks in the past. We conclude 
that more data are badly needed at moderate redshift in the $M_B\sim -18$ 
range if we are to develop a full understanding of the Tully-Fisher relation.

\appendix

\section{The effect of dust corrections on the model predictions}
\begin{figure}
\includegraphics[width=3.5in]{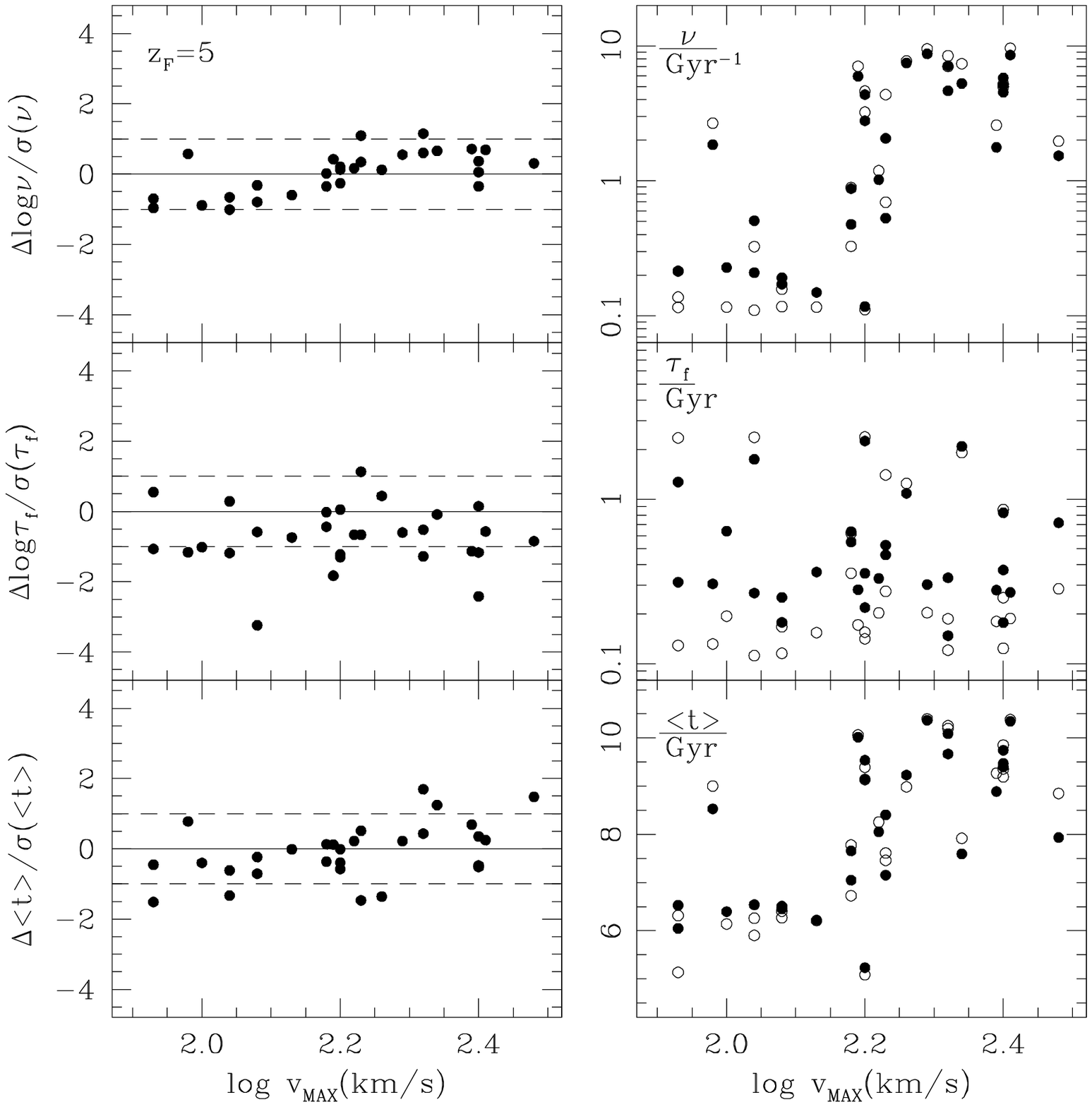}
\caption{Effect of a different dust correction in the model predictions
for the star formation efficiency ({\sl top}), infall timescale ({\sl middle})
and average stellar age ({\sl bottom}) as a function of $\vrot$.
A single formation redshift was chosen ($z_F=5$) with two different
extinction corrections. The left panels show the variation in 
the model predictions with respect to the $1\sigma$ uncertainty. The right
panels give separately the model predictions for these two corrections, 
namely: Tully \& Fouqu\'e (1985; solid dots) and Tully et al. 
(1998; hollow dots). }
\label{fig:dust}
\end{figure}

Dust correction introduces colour variations which can have an
important effect in the model predictions. In this paper we have
used the dust corrections presented in B\"ohm et al. (2004) which 
follow Tully \& Fouqu\'e (1985), so that the correction depends on 
the inclination. However, a systematic effect could arise if this
dust correction were dependent on $\vrot$ as presented in 
Tully et al. (1998). These authors find that luminous disks could
suffer more reddening than low-mass disks, thereby introducing a
systematic trend towards older populations in massive disks. In this
appendix we present the results for a single grid of star formation
histories choosing a formation redshift $z_F=5$ and two extinction
corrections: the one used in this paper -- only dependent on 
inclination -- and the one from Tully et al. (1998) -- where both
the inclination and rotation velocity determine the reddening.
For the latter we also assume no extinction in face-on galaxies.
Figure~\ref{fig:dust} shows the model predictions for the star
formation efficiency ({\sl top}), infall timescale ({\sl middle})
and average stellar ages ({\sl bottom}) as a function of $\vrot$. The
left panels show the change of these parameters with respect to
the uncertainty, and the right panels give the parameters themselves
for these two extinction laws. One can see that the 
parameters vary mostly within the $1\sigma$ uncertainties, 
preserving the strong bimodal trend in star formation efficiency.
No significant systematic trends were seen either by 
B\"ohm et al. (2004) when comparing these two dust correction 
prescriptions. Hence, within the model uncertainties already
discussed in the paper, we conclude that the trend seen with respect
to star formation efficiency or age is unaffected by different dust
correction prescriptions.

\section*{Acknowledgments}
This research has been supported in part by PPARC Theoretical Cosmology
Rolling Grant PPA/G/O/2001/00016 (IF), by the Volkswagen Foundation
(I/76520), and by the ``Deutsches Zentrum f\"ur Luft- und Raumfahrt''
(50\ OR\ 0301). We acknowledge receipt of a JREI grant from PPARC 
to provide the computing facilities. We would like to thank 
Frank Van den Bosch for useful discussions. The anonymous referee is
gratefully acknowledged for a very constructive criticism of this paper.

\bsp

\label{lastpage}

\end{document}